\documentstyle[12pt,epsfig]{article}

\begin{document}

\title{\LARGE \bf Quasi-SU(3) truncation scheme for odd-even $sd$-shell
nuclei.}

\author{
C. E. Vargas$^1$\thanks{Electronic address: cvargas@fis.cinvestav.mx; fellow of
CONACyT.},
J. G. Hirsch$^2$\thanks{Electronic address: hirsch@nuclecu.unam.mx},
J. P. Draayer$^3$\thanks{Electronic address: draayer@lsu.edu},\\
{\small $^1$  Departamento de F\'{\i}sica, Centro de Investigaci\'on y de
Estudios Avanzados del IPN,}\\
{\small Apartado Postal 14-740 M\'exico 07000 DF, M\'exico}\\
{\small $^2$  Instituto de Ciencias Nucleares, Universidad Nacional
Aut\'onoma de M\'exico,}\\
{\small Apartado Postal 70-543 M\'exico 04510 DF, M\'exico }\\
{\small $^3$  Departament of Physics and Astronomy, Louisiana State
University,}\\
{\small Baton Rouge, LA 70803-4001, USA }
}

\date{\today}

\maketitle

\begin{abstract}
{\bf Abstract} The quasi-SU(3) symmetry, as found in shell model calculations
\cite{Zuk94}, refers to the dominance of the single particle plus quadrupole -
quadrupole terms in the Hamiltonian used to describe well deformed 
nuclei, and to the
subspace relevant in its diagonalization. It provides a very 
efficient basis truncation
scheme. It is shown that a small number of SU(3) coupled irreps, 
those with the largest
$C_2$ values within the direct product of the proton and neutron 
SU(3) irreps with spin
0 and 1 (for even number of particles), and spin 1/2 and 3/2 for (for 
odd number of
nucleons), are enough to describe the low energy spectra and B(E2) 
transition strengths
of $^{21}${Ne}, $^{23}${Na} and $^{25}${Mg}. A simple but realistic 
Hamiltonian is
employed. Results compare favorably both with experimental data and 
with full shell
model calculations. Limitations and possible improvements of the 
schematic Hamiltonian
are discussed.

\end{abstract}

\vskip2pc

\section{Introduction.}

Since its introduction more than fifty years ago \cite{May49}, the 
shell model has been
a fundamental tool in the microscopic description of nuclear 
properties. Full shell
model calculations in the $sd$- \cite{Wil84} and $fp$-shells 
\cite{Zuk94} provide very
accurate predictions for energy levels, electromagnetic transition 
strengths and weak
decay half lives. State of the art codes allow studies up to A = 50 
\cite{Sch00}.
The system symmetries, although in general are not exact, provide a natural
truncation scheme of the Hilbert space while keeping the predictive 
power of the
theory, and provide a qualitative understanding of collective nuclear modes.
Algebraic models are particularly well suited to describe systems 
with symmetries,
using either bosonic representations \cite{IBM}, fermionic representations
\cite{Ell58,Hec69}, or combination of both \cite{Isa99}.

In the present article we concentrate our attention in the fermionic 
SU(3) algebraic
model developed by Elliott~\cite{Ell58}. Based on the crucial role 
the quadrupole -
quadrupole interaction plays in deformed systems, the SU(3) algebra is the
natural language to describe quadrupole excitations in a harmonic 
oscillator basis in
light nuclei. The pseudo SU(3) model \cite{Hec69}, built over the pseudo spin
symmetry, plays a similar role in heavy deformed nuclei. Used as an approximate
symmetry, i.e. allowing the mixing of different irreducible representations
(irreps) through the single particle energies and the  pairing 
interaction, it provides
a very good description of low energy bands, B(E2) and B(M1) 
transition strengths in
the rare earth region, both for even-even \cite{Beu00} and 
A-odd~\cite{Var00} deformed
nuclei.

In the pseudo SU(3) model the truncation of the Hilbert space usually 
excludes intruder
orbits, which are known to provide an important contribution to the 
total quadrupole
moment. The quasi SU(3) truncation scheme \cite{Zuk94} offers a 
simple and consistent
way to include intruder orbits in the SU(3) description of heavy 
deformed nuclei. In
\cite{Var98} it was shown that including the leading SU(3) irreps 
(those with the
largest $C_2$ values) and spin 0 and 1 the interplay between the quadrupole -
quadrupole interaction and the spin - orbit splitting is well 
described. A detailed
description of four even-even nuclei in the $sd$-shell, ranging from 
$^{20}$Ne to
$^{28}$Si was presented in \cite{Var000}. Three odd-odd nuclei in the 
$sd$-shell are
discussed in \cite{odd-odd}. The present article deals with the 
odd-mass nuclei in the
$sd$-shell. In the three cases the main goal is to prove that in well 
deformed nuclei,
even in the presence of a large spin-orbit splitting, it is possible 
to make a good
description of the low energy spectra using a small number of SU(3) 
irreps. This is the
meaning of the ``quasi SU(3) truncation scheme". The results 
presented here and in the
two accompanying articles strongly support this conclusion. They open 
the possibility
for a coherent description of nucleons in normal and intruder levels 
in heavy deformed
nuclei using the SU(3) formalism, as envisioned in \cite{Zuk94}.

In the present paper the low energy spectra and B(E2) transition strengths of
$^{21}${Ne}, $^{23}${Na} and $^{25}${Mg} are studied using the quasi 
SU(3) basis and a
schematic Hamiltonian. The band structure is recovered and the wave 
functions along
each band are presented. Results are compared both with the 
experimental data and with
full shell model calculations \cite{Wil84}.
Section 2 reviews the essentials of the SU(3) model, including a 
short description of
the SU(3) basis and the Hamiltonian. In sections 3, 4, and 5 the 
energy spectra, band
structure and B(E2) transition strengths of $^{21}${Ne}, $^{23}${Na} 
and $^{25}${Mg},
respectively, are presented. Section 6 contains the conclusions.

\section{The SU(3) model}

For a general review of the SU(3) model and the operator expansion in terms of
SU(3) tensors we refer the reader to references
\cite{Ell58,Mosh67,Cas87,Cas93,Tro96}, as well as to the first 
article in this series
\cite{Var000}. In what follows we introduce the SU(3) basis and the schematic
Hamiltonian used in this work.

The basis states are written as
\begin{eqnarray}
| \{ n_\pi [f_\pi] \alpha_\pi
(\lambda_\pi,\mu_\pi), n_\nu [f_\nu] \alpha_\nu (\lambda_\nu,\mu_\nu) \} \rho
(\lambda,\mu) k L \{S_\pi,S_\nu\} S ; JM \rangle \label{states}
\end{eqnarray}
where $n_\pi $ is the number of valence protons in the $sd$-shell and 
$[f_\pi] $ is
the irrep of the U(2) spin group for protons, which is associated 
with the spin $\
S_\pi = (f^1_\pi - f^2_\pi)/2$. The SU(3) irrep for protons is 
$(\lambda_\pi,\mu_\pi)$
with a multiplicity label $\alpha_\pi$ associated with the reduction 
from $U(6)$.
Similar definitions hold for the neutrons, labeled with $\nu$. There 
are other two
multiplicity labels: $\rho$, which counts how many times the total 
irrep $(\lambda,
\mu)$ occurs in the direct product $(\lambda_{\pi}, \mu_{\pi}) 
\otimes (\lambda_{\nu},
\mu_{\nu})$ and $K$, which classifies the different occurrences of 
the orbital angular
momentum $L$ in $ (\lambda, \mu)$.

The vector states (\ref{states}) span the complete shell-model space 
within only one
active (harmonic oscillator) shell for each kind of nucleon. As an 
example we take
$^{21}$Ne. It has two protons ($n_\pi = 2$) in the $sd$-shell, which can be
accommodated in three possible irreps:
$(\lambda_{\pi}, \mu_{\pi})=$ (4,0), (2,1) and (0,2). The first and 
third irreps have
spin zero, the second one has spin 1. Each one occurs only once 
($\alpha_\pi = 1$).
For three neutrons in the $sd$-shell there are five irreps:
$(\lambda_{\nu}, \mu_{\nu})=$ (4,1), (2,2), (3,0), (0,3) and (1,1). 
Three of them have
spin 1/2, the other two have spin 3/2. The SU(3) irreps are ordered 
by decreasing
values of the expectation value of the second order Casimir operator, $C_2$,
\begin{equation}
\langle (\lambda,\mu) | C_2 | (\lambda,\mu) \rangle = (\lambda + \mu
+3) ~ (\lambda + \mu) - \lambda \mu .
\end{equation}
Proton and neutron irreps in $^{21}$Ne are listed in Table 1 with 
their spin and $C_2$
values.

\begin{table}
\begin{tabular}{cc|cc}
$(\lambda_\pi, \mu_\pi )S_\pi$ & $C_2$ & $(\lambda_\nu, \mu_\nu 
)S_\nu$ & $C_2$ \\
\hline
(4,0)0  & 28 &(4,1)1/2  & 36 \\
(2,1)1  & 16 &(2,2)1/2  & 24 \\
(0,2)0  & 10 &(3,0)3/2  & 18 \\
         &    &(0,3)3/2  & 18 \\
	&    &(1,1)1/2  &  9
\label{t1}\end{tabular}
\caption{ Irreps, spins and $C_2$ values  for protons and neutrons in
$^{21}${Ne}.}
\end{table}

Calculations performed using the full proton-neutron coupled SU(3) 
Hilbert space in the
$sd$-shell~\cite{Var98,Var000} show that the quasi SU(3) \cite{Zuk94} 
truncation scheme
is quite efficient. The SU(3) basis is built by taking the direct 
product of the proton
and neutron irreps with the largest $C_2$ values and S = 0 and 1 (for even
number of nucleons) or 1/2 and 3/2 (for odd number), and keeping from 
this list only
those states with the largest total $C_2$ values \cite{Var98,Var000,Var99}.
The proton and neutron representations of each studied
nuclei are shown in Tables 1, 5, and 8, and the truncated list of their final
couplings, which describes the Hilbert space used for each nuclei, 
can be seen in
Tables 3, 6, and 9.

The Hamiltonian used is

\begin{eqnarray}
  H & = & H_{sp,\pi} + H_{sp,\nu} - \frac{1}{2}~ \chi~  Q \cdot Q -
	~ G_\pi	~H_{pair,\pi} ~\label{eq:ham} \\
    &   & - ~G_\nu ~H_{pair,\nu} + ~a~ K_J^2~ +~ b~ J^2~ +~ A_{sym}~
          C_2 . \nonumber
\end{eqnarray}

\noindent where $H_{sp,\alpha}$ is the spherical Nilsson Hamiltonian 
for $\alpha$ =
$\pi$ or $\nu$ and the quadrupole-quadrupole and pairing interaction 
strengths $\chi$,
$G_\pi$ and $G_\nu$ has been fixed from systematics as in previous
work~\cite{Var00,Var000}. The parameters $a$, $b$, and $A_{sym}$ 
correspond to the
three ``rotor terms''. They are small, and provide some freedom to 
perform a nuclei by
nuclei best fit. The Hamiltonian parameters used in this work are 
listed in Table 2.

\begin{table}
\begin{tabular}{c|ccccc}
Nucleus & $\chi$  & $G_\pi = G_\nu$ &$a$ & b  & $A_{sym}$ \\ \hline
$^{21}${Ne} & 0.1063& 0.4524 & -0.200 &-0.020 &  0.010 \\
$^{23}${Na} & 0.0914& 0.4130 &    0   &-0.005 &    0   \\
$^{25}${Mg} & 0.0795& 0.3800 & -0.130 & 0.090 & 0.024
\label{t2}\end{tabular}
\caption{ Hamiltonian parameters in MeV.}
\end{table}

For the spherical Nilsson Hamiltonian
\begin{eqnarray}
H_{sp,\alpha} & = & \hbar \omega_o \{ ( \hat{\eta} + \frac{3}{2} ) - 
2 \kappa ~ \vec{l}
\cdot \vec{s} - \kappa \mu ~ {\vec{l}}^2 \}
\label{eq:nilsson}\end{eqnarray}
the  Ring and  Schuck~\cite{Rin79} parametrization $\kappa = 0.08$ 
and $\mu = 0.000$ was chosen.
While both the single
particle Hamiltonian and the pairing terms induce a mixing of SU(3) 
irreps, for the
values used here the single particle terms are by far and away the 
driving force behind
the mixing.

Hamiltonian (\ref{eq:ham}) proved to be very powerful in the 
description of the normal
parity bands in heavy deformed A-odd nuclei~\cite{Var00},  where 
protons and neutrons
occupy different major shells. In the description of light nuclei, 
the same Hamiltonian
is missing the proton-neutron pairing term, and for this reason it is 
not isoscalar.
Given that the main goal of the present work is to assert the 
validity of the quasi
SU(3) truncation scheme, we have kept this Hamiltonian, using its rotor terms
to partially compensate for the missing terms. Further applications 
of this model to
light and medium mass nuclei must use an isospin invariant 
Hamiltonian if the model is
intended to display its full predictive power.

In the following sections a detailed analysis of the nuclei 
$^{21}${Ne}, $^{23}${Na}
and $^{25}${Mg} is presented. The energy spectra, band structure and 
B(E2) transition
strengths are shown in each case.

\section{$^{21}${Ne}}

$^{21}${Ne} has two protons and three neutrons occupying the  $sd$ 
valence shell.
As discussed in the previous section, they can be in any of the three 
SU(3) proton
irreps and five SU(3) neutron irreps listed in Table 1, which include 
proton states
with spin 0 and 1, and neutron states with spin 1/2 and 3/2. The truncated
proton-neutron coupled basis is listed in Table 3. It includes the 17 
coupled irreps
with the largest $C_2$ values.

\begin{figure}
\vspace*{-2.5cm}
\epsfxsize=16cm
\centerline{\epsfbox{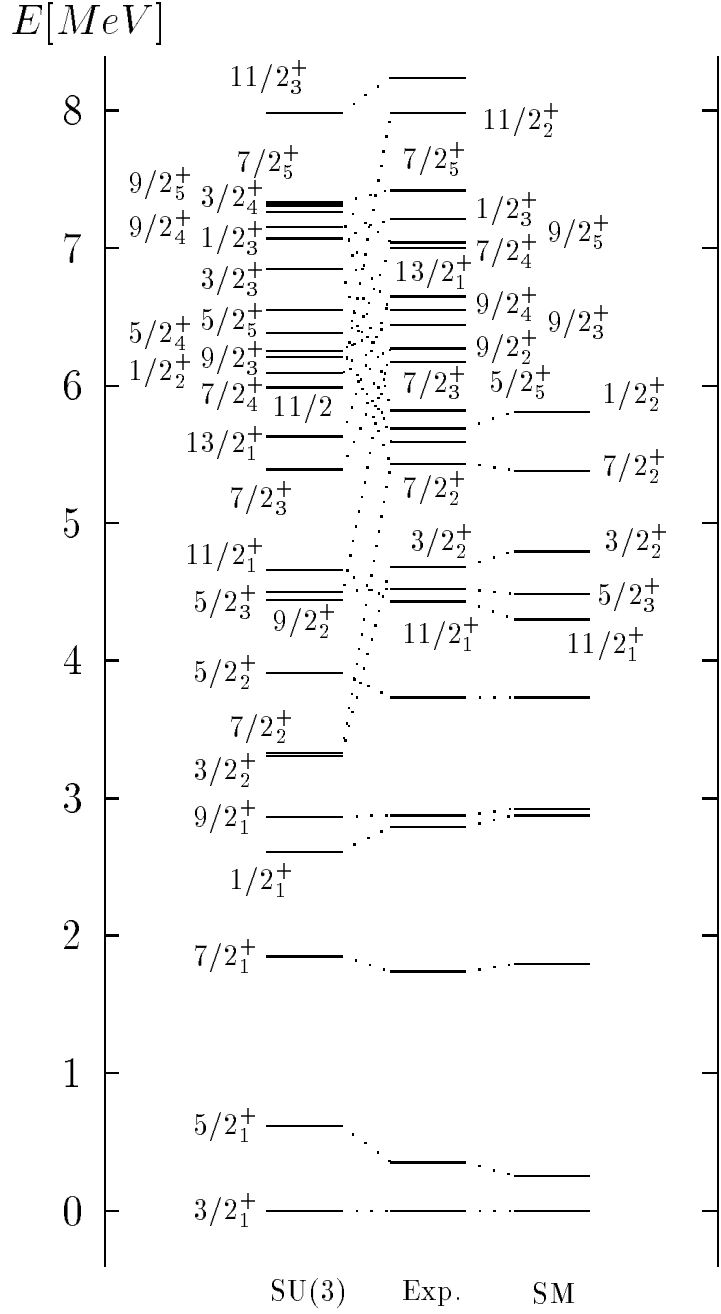}}\vspace*{-1.2cm}
\caption{ Energy spectrum for $^{21}${Ne}. The first column displays
the present results, the second shows experimental data, and the third
column display full shell-model results \cite{Hof89} with a unified $sd$
Hamiltonian \cite{Wil84}.}
\label{fig1}\end{figure}

\begin{figure}
\epsfxsize=10.5cm
\centerline{\epsfbox{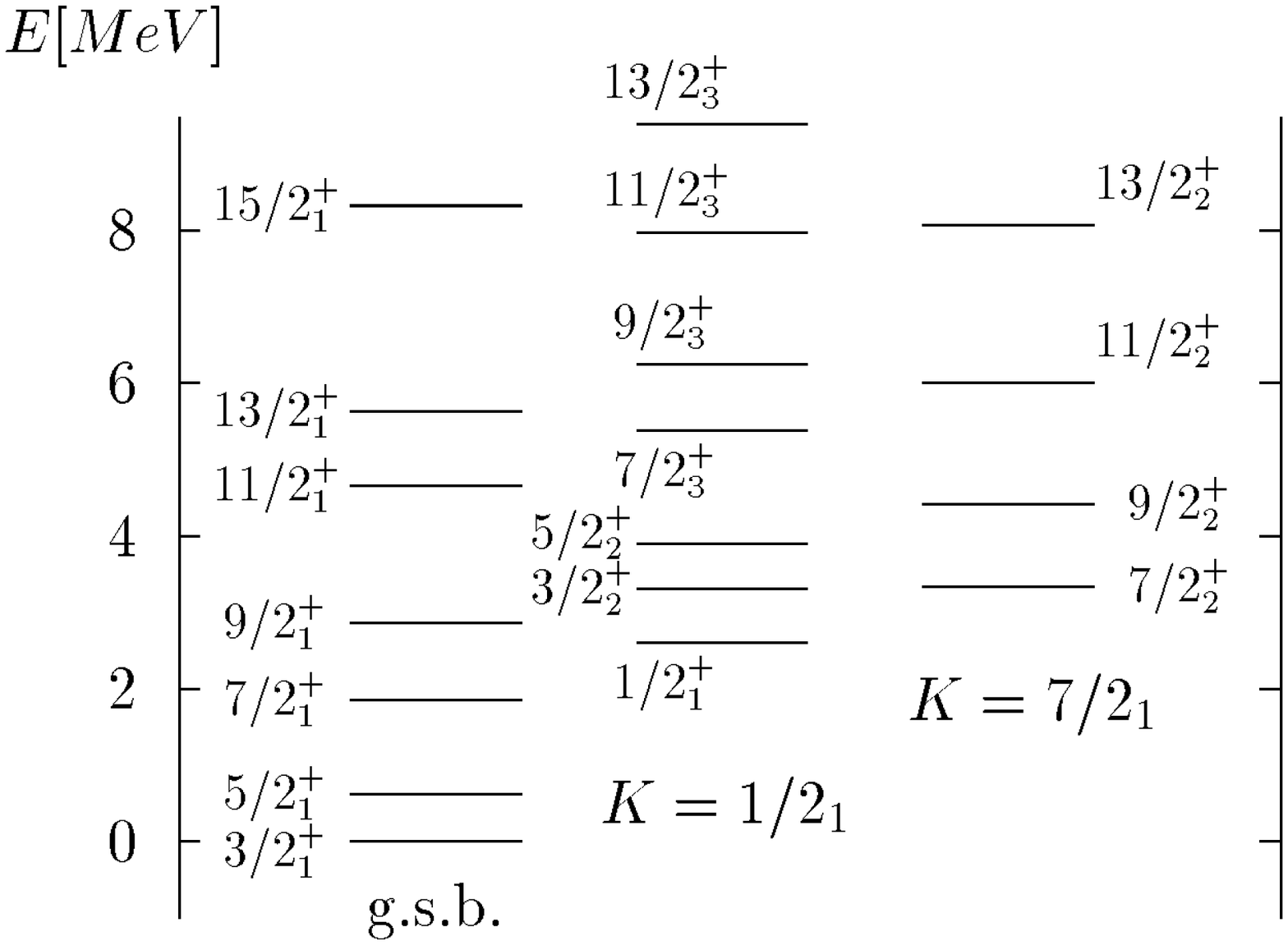}}\vspace*{-4.7cm}
\caption{ Band structure in $^{21}${Ne}.}
\label{fig2} \end{figure}

\begin{figure}
\vspace*{-3.2cm}
\epsfxsize=15.5cm
\centerline{\epsfbox{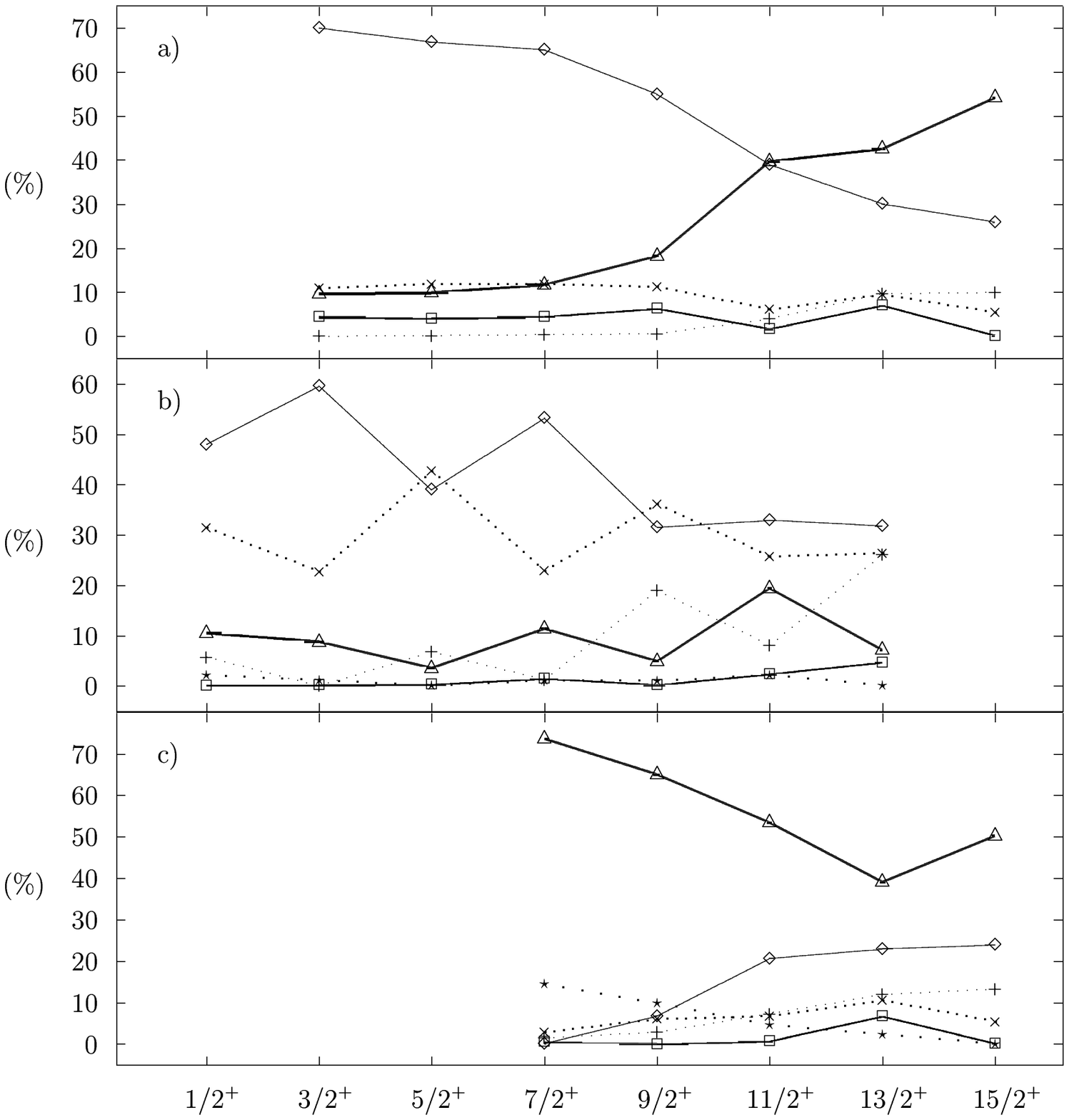}}\vspace*{-3.7cm}
\caption{ Wave function components of states belonging to a) the ground
state band, b) the K = 1/2 band and c) the K = 7/2 band in $^{21}$Ne. The
percentage each irrep contributes is shown as function of the angular
momentum. The convention used is $\Diamond$ for (8,1)1/2[(4,0)0
$\otimes$ (4,1)1/2], $+$ for (4,3)3/2[(2,1)1 $\otimes$ (2,2)1/2], $\Box$
for (7,0)3/2[(4,0)0 $\otimes$ (3,0)3/2], $\times$ for (6,2)1/2[(4,0)0
$\otimes$ (2,2)1/2], $\triangle$ for (6,2)3/2[(2,1)1 $\otimes$ (4,1)1/2]
and $\star$ for (4,3)1/2[(2,1)1 $\otimes$ (2,2)1/2].}
\label{fig3} \end{figure}

The $^{21}${Ne} energy spectra obtained with this basis using the
Hamiltonian parameters shown in the first row of Table 2 are presented in
the right hand side column of Fig. \ref{fig1}. They are compared  with
the experimental results \cite{nndc}, shown in the second column, and with
the theoretical results obtained with full shell model calculations using
the unified s - d shell Hamiltonian \cite{Hof89} (third column).
We also show the band structure in Fig. \ref{fig2} and their wave
function components  in Fig. \ref{fig3}.

\begin{table}
\begin{tabular}{cc|llll}
$(\lambda_\pi, \mu_\pi )S_\pi$ & $(\lambda_\nu, \mu_\nu )S_\nu$ &
\multicolumn{4}{c}{$(\lambda, \mu )S$ total} \\ \hline
(4,0)0& (4,1)1/2& (8,1)1/2 & (6,2)1/2 & (7,0)1/2 & (4,3)1/2 \\
(4,0)0& (2,2)1/2& (6,2)1/2 & (4,3)1/2 \\
(2,1)1& (4,1)1/2& (6,2)1/2,3/2 & (7,0)1/2,3/2 & (4,3)1/2,3/2 \\
(4,0)0& (3,0)3/2& (7,0)3/2  \\
(4,0)0& (0,3)3/2& (4,3)3/2  \\
(2,1)1& (2,2)1/2& (4,3)1/2,3/2  \\
(0,2)0& (4,1)1/2& (4,3)1/2
\label{t3}\end{tabular}
\caption{ The 17 irreps used in description of $^{21}${Ne}. In some
couplings the total irrep can have spins of 1/2 and/or 3/2.}
\end{table}

It can be seen
from Fig. \ref{fig1} that the  low-lying energy spectra of $^{21}${Ne}
is well described using the truncated quasi-SU(3) space,
which
includes only
the 17 coupled irreps with the largest $C_2$ values, out of the 137
coupled irreps (some of them with $\rho$ multiplicity larger than
1) which form the full SU(3) basis.

Despite the very schematic form of Hamiltonian (\ref{eq:ham}) and the
missing proton-neutron terms mentioned above, the rotor terms allow a fine
tuning of the spectra. The moment of inertia is slightly increased by
selecting b = -0.02. It corrects the quadrupole moment of inertia $\sc I$,
where $ 1/ (2 {\sc I}) = (3/2) \chi \approx 0.16$. Some excited states are
moved to higher energies by using $a$ = -0.200. Using $A_{sym}$ = 0.010
for the symmetry term enhances the contributions of the irreps with
$\lambda$ and $\mu$ even relative to the others, because they belong to
different symmetry types of the intrinsic Vierergruppe $D_2$ \cite{Les87}.

It is worth mentioning that, as it was the case in even-even nuclei
\cite{Var000}, the main features of the spectra can be obtained by setting
all the rotor parameters to zero. On the other hand, if they are values
that are too large (in absolute value), the whole band structure is destroyed,
and with it the agreement with the observed B(E2) values (see below).

The present model predicts the energies of
the ground state band, and the states $1/2_1$, $5/2_2$, $5/2_3$, $11/2_3$,
$9/2_{3,6}$, $7/2_5$, and $1/2_2$ pretty close to their experimental
counterparts. The ground state band staggering effect forces the states to
be clustered by pairs: (3/2, 5/2), (7/2, 9/2), (11/2, 13/2) (see also
Fig. 2). It is slightly exaggerated in the last case. The levels
$3/2_2$, $7/2_2$, $9/2_2$, $7/2_4$, and $11/2_2$ are predicted at energies
around 2 MeV lower than observed. This is a clear limitation of the model,
which can be related to the limited and schematic nature of Hamiltonian. 
Further
investigations including proton-neutron pairing would clarify this 
point. In general, the
results reported here represent a clear improvement from previous SU(3) based
descriptions \cite{Naq92}.

\begin{table}
\begin{tabular}{c|ccc}
       $B(E2) \uparrow $       & Expt.  & SU(3) &  SM \cite{Hof89} \\ \hline
$3/2_1^+ \rightarrow 5/2_1^+$ & $1.239 \pm 0.155$ & 1.215 & 1.126 \\
$5/2_1^+ \rightarrow 7/2_1^+$ & $0.505 \pm 0.183$ & 0.710 & 0.738 \\
$7/2_1^+ \rightarrow 9/2_1^+$ & $0.387 \pm 0.215$ & 0.348 & 0.391 \\
$9/2_1^+ \rightarrow 11/2_1^+$& $0.248 \pm 0.165$ & 0.239 & 0.247 \\
$11/2_1^+ \rightarrow 13/2_1^+$&                  & 0.187 & 0.136 \\
$13/2_1^+ \rightarrow 15/2_1^+$&                  & 0.129 & 0.078 \\
$15/2_1^+ \rightarrow 17/2_1^+$&                  & 0.077 & 0.077 \\
$17/2_1^+ \rightarrow 19/2_1^+$&                  & 0.040 & 0.034 \\
$3/2_1^+ \rightarrow 7/2_1^+$ & $0.675 \pm 0.151$ & 0.688 &Ý0.633 \\
$5/2_1^+ \rightarrow 9/2_1^+$ & $0.906 \pm 0.097$ & 0.771 & 0.728 \\
$7/2_1^+ \rightarrow 11/2_1^+$&                   & 0.618 & 0.707 \\
$9/2_1^+ \rightarrow 13/2_1^+$&                   & 0.540 & 0.564 \\
$11/2_1^+ \rightarrow 15/2_1^+$&                  & 0.413 & 0.422 \\
$13/2_1^+ \rightarrow 17/2_1^+$&                  & 0.264 & 0.318 \\
$15/2_1^+ \rightarrow 19/2_1^+$&                  & 0.182 & 0.215 \\
$5/2_1^+ \rightarrow 7/2_4^+$ &  0.011            & 0.002 &
\label{t4}\end{tabular}
\caption{ B(E2) transition strengths for $^{21}${Ne} in [$e^2b^2 \times
10^{-2}$]. }
\end{table}

Effective charges $e_{\pi~ eff} = 1.56 e, e_{\nu ~eff} = 0.56 $ were used
in the evaluation of  B(E2) transition strengths for the three nuclei.
  B(E2) values
calculated
with the present model, those obtained by the shell model \cite{Hof89}
and experimental ones for $^{21}${Ne} are shown in Table 4.
The agreement
with both the experimental and shell model values
is in general very good. All but the last transition connect
states belonging to the ground state band. The factor 5 deviation 
found in the $5/2_1^+
\rightarrow 7/2_4^+$ transition seem to reflect effects of the truncation of
the Hilbert space.

The $^{21}$Ne band structure of this nuclei is presented in Fig. \ref{fig2}.
Results for the the ground state and K = 1/2 bands are very similar to those
found in previous studies  \cite{Hof89}. The staggering in
the ground state band noted above is clearly seen.

The SU(3) wave function components of each band are shown in  Fig.
\ref{fig3}. The percentage different SU(3) irreps contribute to each
state is plotted as a function of the angular momentum of the members of
the band. Different irreps are  represented by different lines and symbols
(listed in the figure caption). All irreps which contributes more than 2\%
are plotted. The bands are recognized by their large B(E2) transition
strengths. The slow (adiabatic) change in their SU(3) content helps to
confirm the band assignment.

Insert a) of  Fig. \ref{fig3} shows the dominance of the irrep (8,1) with
spin 1/2 in the ground state band up to J=9/2, in agreement with previous
studies \cite{Naq92,Fel77}. At J = 11/2 there is a clear change in the
wave function, which for J=13/2, 15/2 is dominated by the (6,2) irrep with
spin 3/2. Irreps with spin S=3/2 were not included in \cite{Naq92}. In the
present contribution we are showing for the first time its relevance in
the low energy spectra. This is the main new feature in the present quasi
SU(3) truncation scheme. In  \cite{Fel77} it was found that even in
presence of strong SU(3) breaking interactions, like the one proposed by
Preedom and Wildenthal, there is a clear dominance of the (8,1)1/2 irrep
for the ground state, in complete agreement with the present results.
In the other two bands, insert b) for K = 1/2 and c) for K = 7/2 in
Fig. \ref{fig3}, similar features can be found. While the K = 1/2 band
exhibits strong mixing between the (8,1) 1/2 and (6,2) 1/2 irreps, the K =
7/2 band is dominated by the (6,2) 3/2 band, underlining once again the
crucial role played by spin 3/2 irreps.

\section{$^{23}${Na}}

$^{23}${Na} has 11 protons and 12 neutrons. The valence space contains 3
protons and 4 neutrons in the $sd$-shell, allowing for the  five proton
SU(3) irreps and ten neutron SU(3)  irreps listed in Table 5. From them,
only three proton irreps and five neutron irreps contribute to the
truncated basis, which include the 20 coupled irreps with spin S= 1/2, 3/2
listed in Table 6. The complete space contains 670 coupled irreps plus
their external multiplicities. Notice that most irreps can have both
spins, due to the $ 1 \otimes 1/2$ coupling. There are also two (7,2) 1/2
irreps, one coming from the (4,1)1/2 $\otimes$(3,1)0 coupling, and the
other from the (4,1)1/2 $\otimes$ (3,1)1 coupling.

\begin{table}
\begin{tabular}{lc|lc}
$(\lambda_\pi, \mu_\pi )S_\pi$ & $C_2$ & $(\lambda_\nu, \mu_\nu )S_\nu$ &
$C_2$ \\ \hline
(4,1)1/2& 36 &(4,2)0  & 46 \\
(2,2)1/2& 24 &(5,0)1  & 40 \\
(3,0)3/2& 18 &(2,3)1  & 34 \\
(0,3)3/2& 18 &(0,4)0  & 28 \\
(1,1)1/2&  9 &(3,1)0,1& 25 \\
         &    &(1,2)1,2& 16 \\
         &    &(2,0)0  & 10 \\
         &    &(0,1)1  &  4
\label{t5}\end{tabular}
\caption{ Irreps, spins and  $C_2$ values for protons and
neutrons in $^{23}${Na}.}
\end{table}

\begin{table}
\begin{tabular}{ll|llcc}
$(\lambda_\pi, \mu_\pi )$ & $(\lambda_\nu, \mu_\nu )$ &
\multicolumn{4}{c}{$(\lambda, \mu )$ total} \\ \hline
(4,1)1/2& (4,2)0 & (8,3)1/2 & (9,1)1/2 & (6,4)1/2 & (7,2)1/2 \\
(4,1)1/2& (5,0)1 & (9,1)1/2,3/2 & (7,2)1/2,3/2 \\
(4,1)1/2& (2,3)1 & (6,4)1/2,3/2 & (7,2)1/2,3/2 \\
(2,2)1/2& (4,2)0 & (6,4)1/2 & (7,2)1/2 \\
(4,1)1/2& (3,1)0,1&(7,2)1/2,3/2 \\
(2,2)1/2& (5,0)1 & (7,2)1/2,3/2 \\
(3,0)3/2& (4,2)0 & (7,2)3/2
\label{t6}\end{tabular}
\caption{ The 20 irreps used in the description of $^{23}${Na}.}
\end{table}

\begin{figure}
\vspace*{-2.3cm}
\epsfxsize=16.5cm
\centerline{\epsfbox{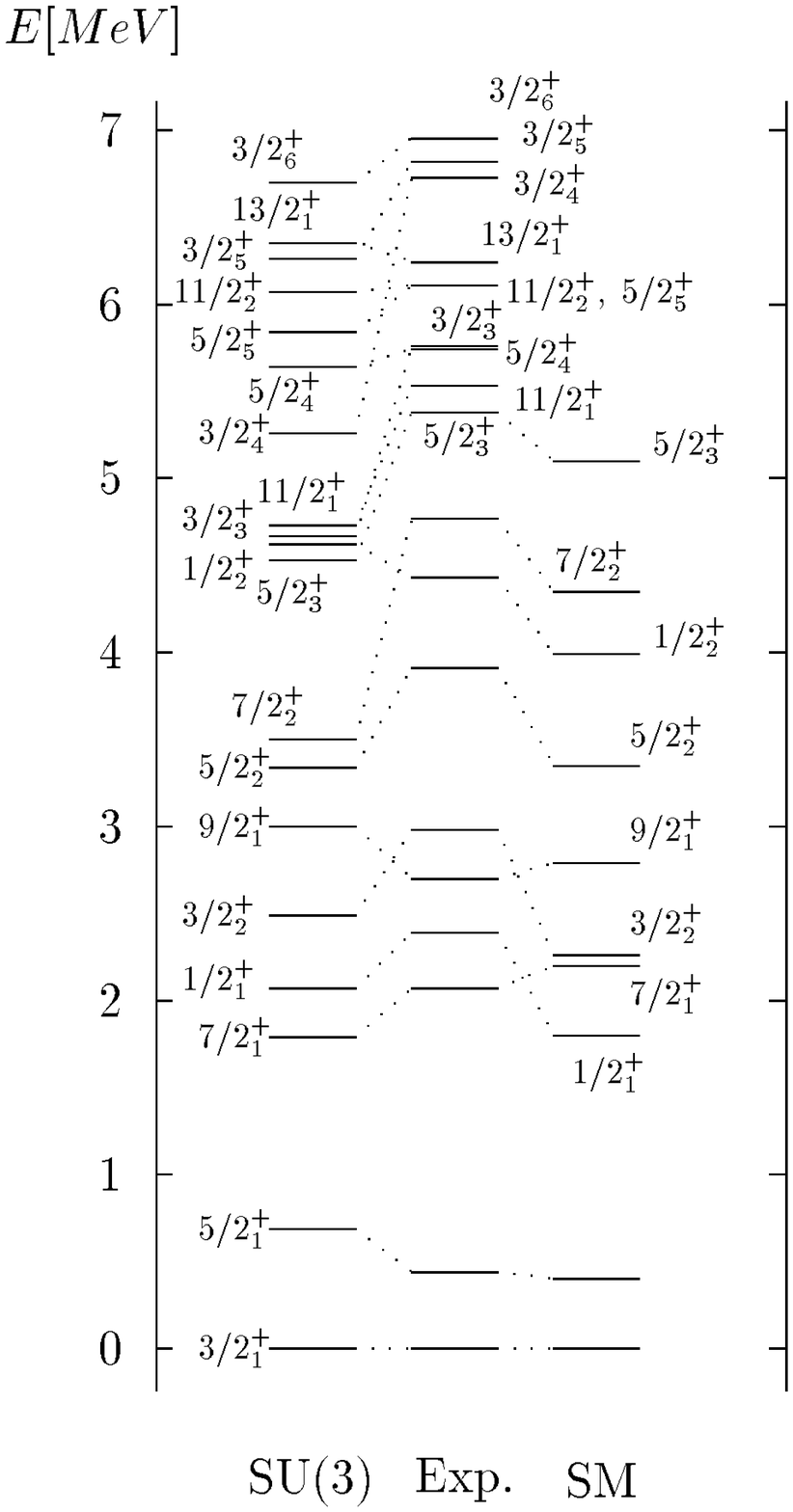}}\vspace*{-1.5cm}
\caption{ Energy spectrum of $^{23}${Na}, with the same convention of
Fig. \ref{fig1}. The SM data are from Glasgow group \cite{Col74}.}
\label{fig4}\end{figure}

The right hand side of Fig. \ref{fig4} shows the results obtained for
the low energy spectra of $^{23}${Na}, calculated using the Hamiltonian
parameters listed in the second row of Table 2, and the Hilbert space
described above. This is compared with the experimental data \cite{nndc},
presented in the second column, and with full shell model calculations
performed by the Glasgow group \cite{Col74} (third column).
The ground state, as well as some excited states are well described using the
quasi SU(3) truncation scheme. On the other hand, as can be seen in Fig.
\ref{fig4}, the model fails to
reproduce the energy of the states $7/2_2$, $5/2_3$, $3/2_3$, $3/2_4$, and
others, thereby exhibiting the limits of the model. The energy of these states
is in general underestimated, probably due to the limited Hamiltonian used.

\begin{figure}
\epsfxsize=15.5cm
\centerline{\epsfbox{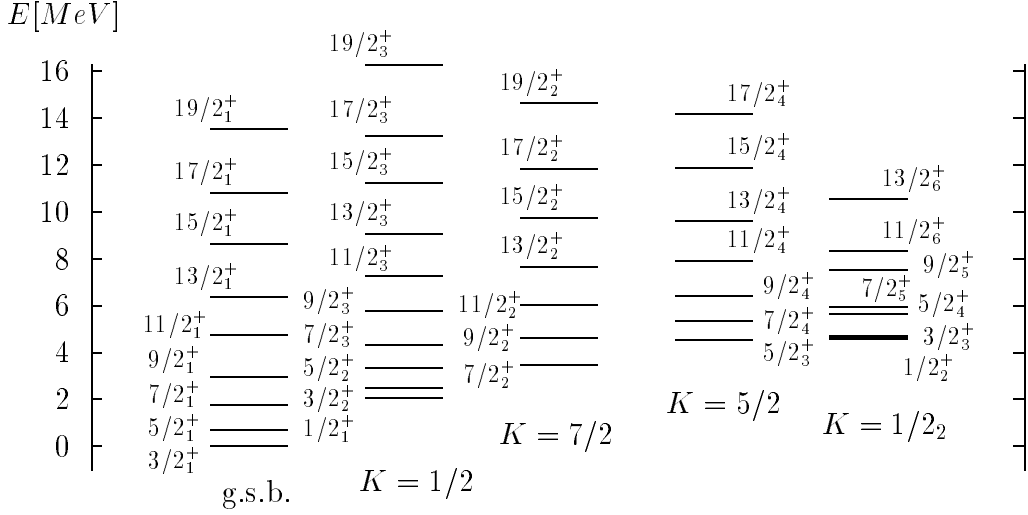}}\vspace*{-11.7cm}
\caption{ Band structure in $^{23}${Na}.}
\label{fig5}\end{figure}

The band structure, reconstructed from the largest B(E2) calculated
values, is shown in Fig. \ref{fig5}. Comparing Figs. \ref{fig4} and
\ref{fig5} it is noticeable that the ground state band is well described,
while the excited K=1/2 and K=7/2 bands are displaced to lower energies
than in the observed spectrum. Adjusting the Hamiltonian parameters 
did not prove to be
useful for correcting this feature, pointing again to the need of
including another terms in the Hamiltonian.

\begin{figure}
\vspace*{-3.7cm}
\epsfxsize=15.5cm
\centerline{\epsfbox{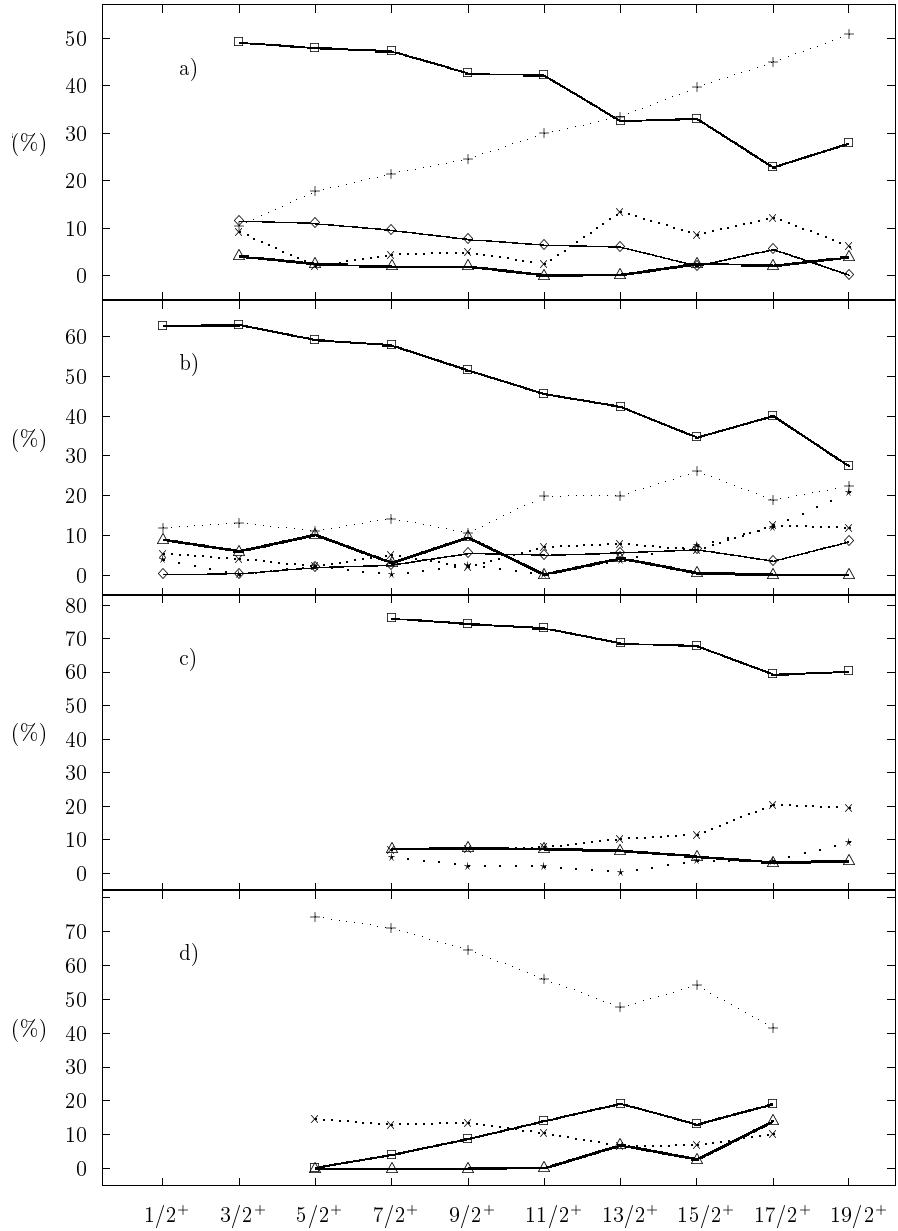}}\vspace*{-3.1cm}
\caption{ Wave function components of states belonging to a) the ground
state band, b) the K = 1/2 band, c) the K = 7/2 band and d) the K = 5/2
band in $^{23}$Na. The percentage associated with each irrep is shown as
function of the angular momentum. The convention used is $\Box$ for
(8,3)1/2[(4,1)1/2 $\otimes$ (4,2)0], $+$ for (9,1)3/2[(4,1)1/2 $\otimes$
(5,0)1], $\Diamond$ for (9,1)1/2[(4,1)1/2 $\otimes$ (4,2)0], $\times$ for
(7,2)3/2[(4,1)1/2 $\otimes$ (3,1)1], $\triangle$ for (6,4)1/2[(2,2)1/2
$\otimes$ (4,2)0] and $\star$ for (6,4)3/2[(4,1)1/2 $\otimes$ (2,3)1].}
\label{fig6}\end{figure}

The negative parameter b = -0.005 provides a small correction to the moment of
inertia, which is similar to what was found for $^{21}$Ne. The other two rotor
parameters were found to be useless in this nuclei. Both tend to wash 
out the staggering
in the ground state band, and also destroy the band structure. For 
these reasons they were
fixed at zero.

It is remarkable that many states are well described,  and that in 
most cases the small
deviations with respect to the experimental values show the same 
trend as the shell model
results. Some small energy displacements from the observed data in 
the ground state
band are similar to those found in previous SU(3) studies \cite{Naq95}.

A study was also performed using the same Hamiltonian but increasing 
the basis for
20 to 302 irreps, around 40\% of the full space. The energy spectrum 
showed almost
no changes. This is a very strong argument in favor of the quasi 
SU(3) truncation scheme.
The most sizable change happened for the $5/2_1$ state, whose energy 
was reduced by 200
keV, and thereby approaching the experimental value.

\begin{table}
\begin{tabular}{c|cc}
$B(E2) \uparrow $             & Expt.             & SU(3) \\ \hline
$3/2_1^+ \rightarrow 5/2_1^+$ & $1.573 \pm 0.233$ & 1.677 \\
$5/2_1^+ \rightarrow 7/2_1^+$ & $0.777 \pm 0.207$ & 1.123 \\
$5/2_1^+ \rightarrow 9/2_1^+$ & $1.359 \pm 0.194$ & 1.031 \\
$3/2_1^+ \rightarrow 7/2_1^+$ & $0.987 \pm 0.085$ & 1.015 \\
$1/2_1^+ \rightarrow 5/2_1^+$ & $0.124 \pm 0.023$ & 0.002 \\
$1/2_1^+ \rightarrow 5/2_2^+$ & $2.681 \pm 0.583$ & 2.103 \\
$7/2_1^+ \rightarrow 9/2_1^+$ &     	  & 0.447Ý\\
$7/2_1^+ \rightarrow 11/2_1^+$ &   	  & 1.216Ý\\
$9/2_1^+ \rightarrow 11/2_1^+$ &  	  & 0.512Ý\\
$9/2_1^+ \rightarrow 13/2_1^+$ &    	  & 0.888Ý\\
$11/2_1^+ \rightarrow 13/2_1^+$ &   	  & 0.157Ý\\
$11/2_1^+ \rightarrow 15/2_1^+$ & 	  & 1.055Ý\\
$13/2_1^+ \rightarrow 15/2_1^+$ &   	  & 0.322Ý\\
$13/2_1^+ \rightarrow 17/2_1^+$ &  	  & 0.671Ý\\
$15/2_1^+ \rightarrow 19/2_1^+$ &  	  & 0.800Ý\\
\label{t7}\end{tabular}
\caption{ B(E2) transitions for $^{23}${Na} in [$e^2b^2 \times
10^{-2}$].}
\end{table}

Fig. \ref{fig6} shows the wave function decomposition of the states 
belonging to a)
the ground state band, b) the K = 1/2 band, c) the K = 7/2 band and 
d) the K = 5/2
band in $^{23}$Na. For each band the percentage associated with 
different SU(3) irreps
is shown as function of the angular momentum. The ground state band 
is dominated by the
irrep (8,3) 1/2 up to J = 13/2. For larger angular momentum the irrep 
(9,1) 3/2 has the
largest component. The K = 1/2 band has a similar structure, but for 
large angular
momentum there is competition between the (9,1) 3/2 and (6,4) 3/2 
irreps to replace
the (8,3) 1/2, resulting in a strong mixing. The K = 7/2 band is 
dominated by the
(8,3) 1/2 irrep, while the K = 5/2 band is by large made of the (9,1) 
3/2 irrep.

It must be stressed again that the inclusion of the  (9,1) 3/2 and other
spin 3/2 irreps is one of the most relevant features of the quasi
SU(3) truncation scheme, which accounts for most of its present success,
as compared with previous SU(3) studies \cite{Naq95}.

B(E2) transition strengths are presented in Table 7. They are  compared
with the experimental values, available for some transitions between
states belonging to the ground state band and to the excited K = 1/2
band. The agreement is in general very good.
The calculated B(E2;$1/2_1 \rightarrow 5/2_1$) value is two orders of magnitud
smaller than the experimental one, while B(E2;$1/2_1 \rightarrow 5/2_2$) is
larger and well reproduced.

\section{$^{25}${Mg}}

$^{25}${Mg} has 4 protons and 5 neutrons in the $sd$-valence shell. The
ten proton
and 12 neutron irreps in which they can be accommodated
are listed in Table 8. From the 1821 proton-neutron coupled SU(3) irreps
we selected the 16 with the largest $C_2$ values to build the truncated
Hilbert space. They are shown in Table 8, together with their spin.  As in
the previous cases, some coupled irreps can have both spin 1/2 and
3/2. Only the three proton SU(3) irreps and the four neutron SU(3) irreps
with the largest $C_2$ values were included in the truncated basis.

\begin{figure}
\vspace*{-1cm}
\epsfxsize=15.5cm
\centerline{\epsfbox{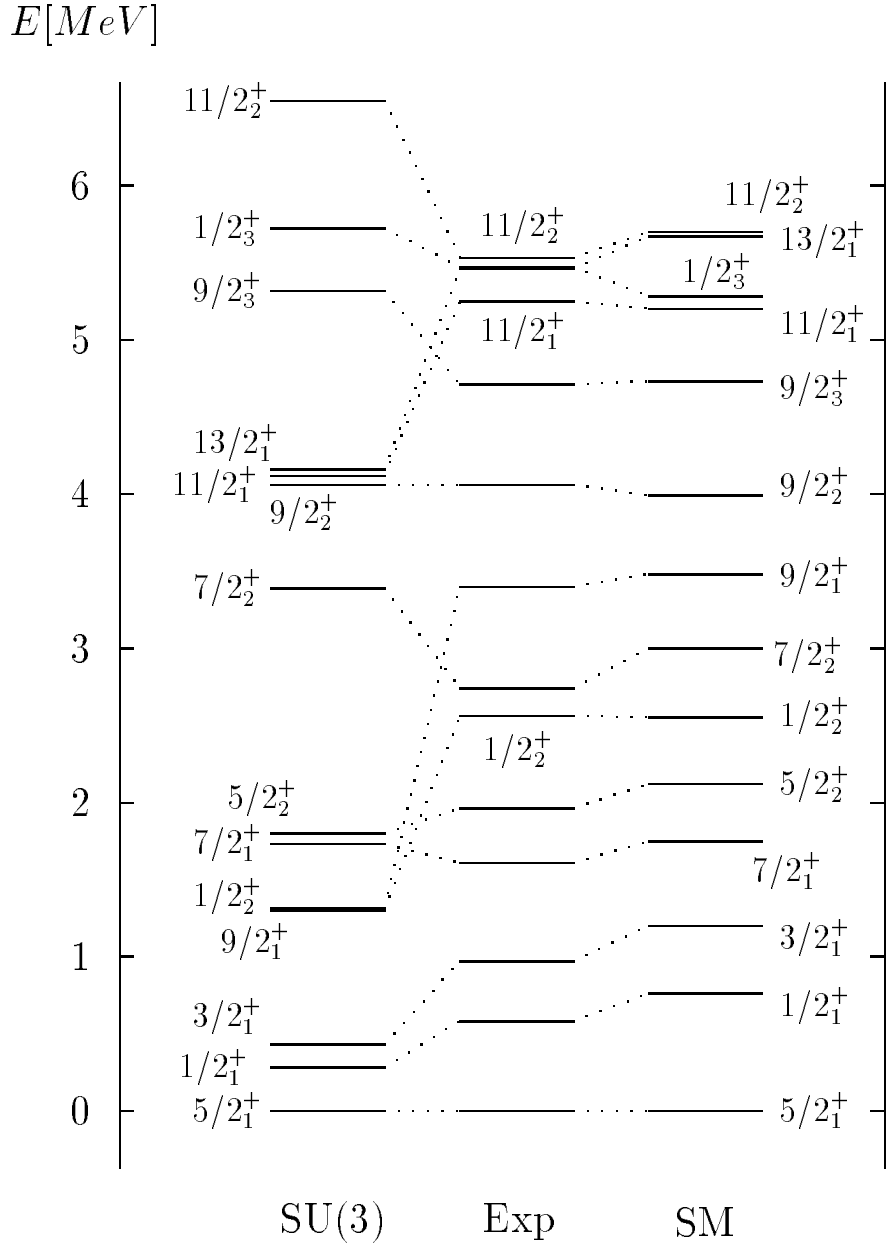}}\vspace*{-1.5cm}
\caption{ Energy spectra of $^{25}${Mg}. The first column shows
the SU(3) results, the second the experimental levels, and the third
the results of full shell model calculations \cite{Hea88}.}
\label{fig7}\end{figure}

\begin{figure}
\vspace*{-1.7cm}
\epsfxsize=15.5cm
\centerline{\epsfbox{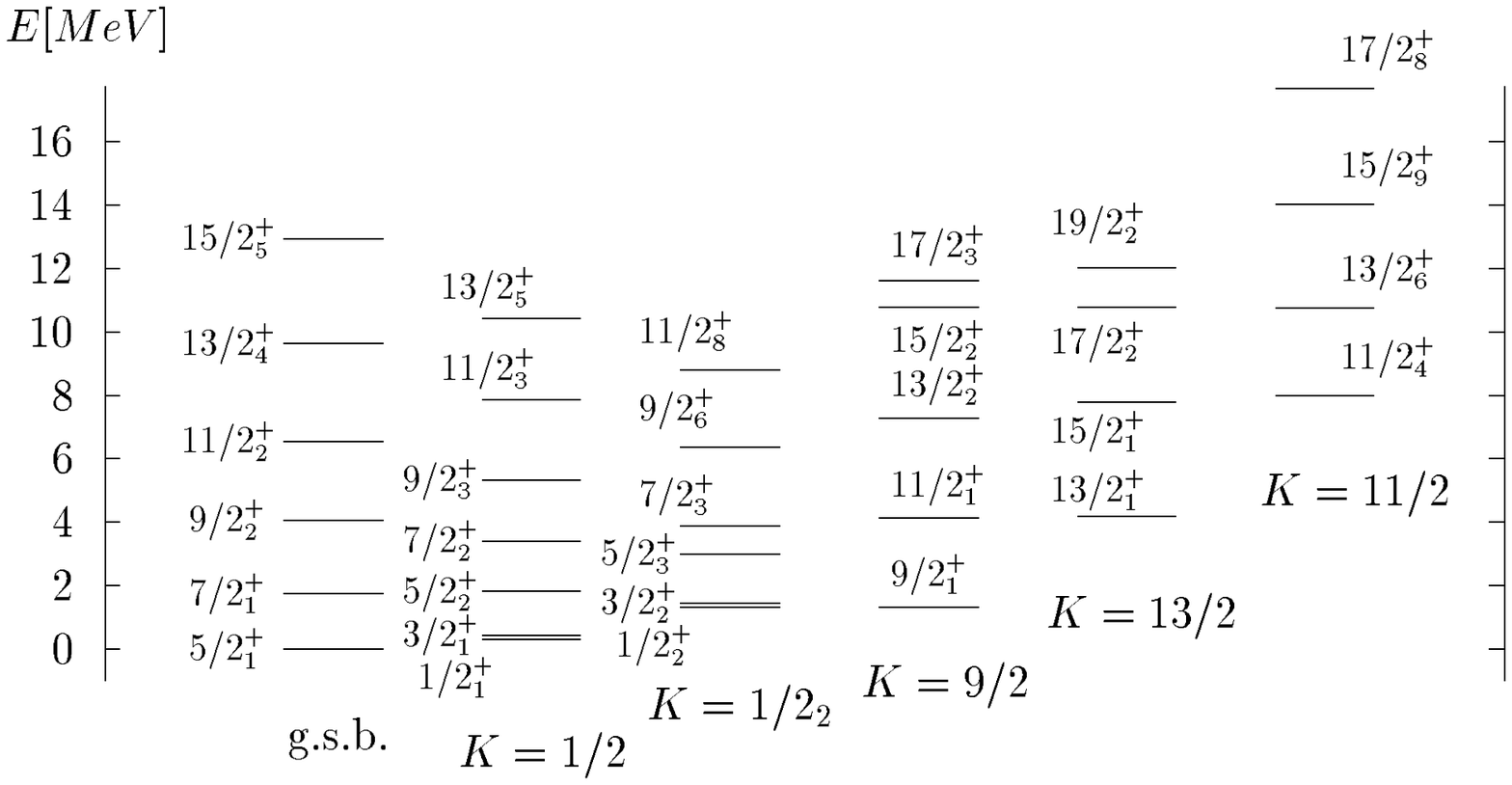}}\vspace*{-10.7cm}
\caption{ Band structure of $^{25}${Mg}.}
\label{fig8}\end{figure}

\begin{figure}
\vspace*{-4.2cm}
\epsfxsize=14cm
\centerline{\epsfbox{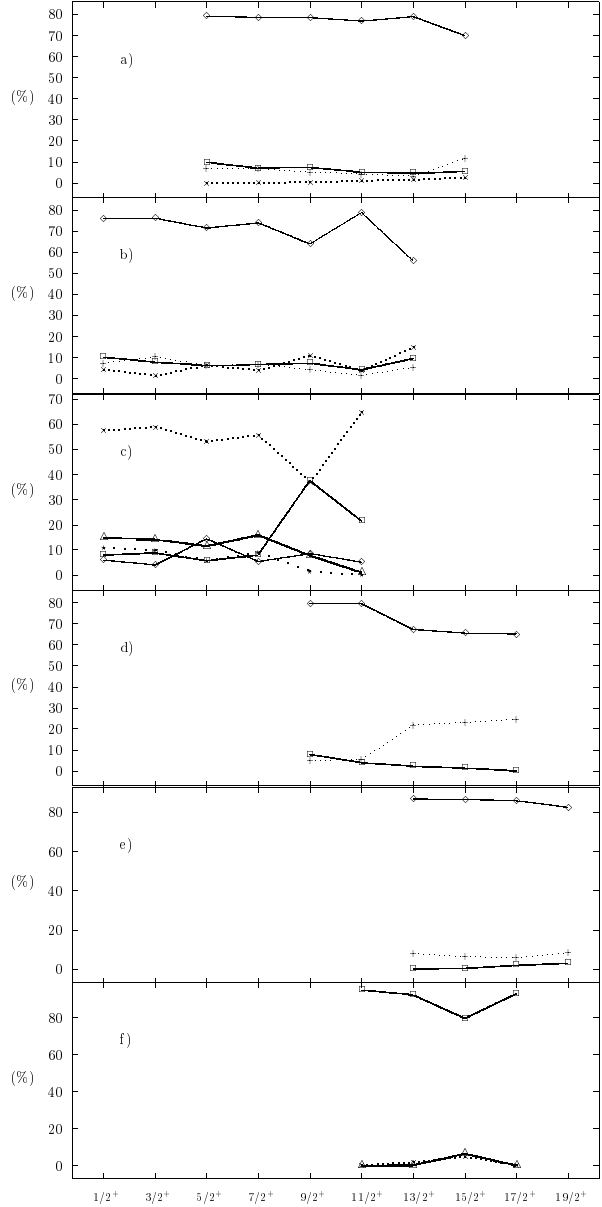}}\vspace*{-1cm}
\caption{ Wave function components of states belonging to a) the ground
state band, b) the K = 1/2 band, c) the K = $1/2_2$ band, d) the K = 9/2
band, e) the K = 13/2 band and f) the K = 11/2 band in $^{25}$Mg. The
percentage each irrep contributes is shown as function of the angular
momentum. The convention used is $\Diamond$ for (6,6)1/2[(4,2)0
$\otimes$ (2,4)1/2], $+$ for (4,7)3/2[(2,3)1 $\otimes$ (2,4)1/2],
$\times$ for (9,3)1/2[(4,2)0 $\otimes$ (5,1)1/2], $\triangle$ for
(10,1)3/2[(5,0)1 $\otimes$ (5,1)1/2], $\star$ for (10,1)1/2[(4,2)0
$\otimes$ (5,1)1] and $\Box$ for (7,4)3/2 $\{$ [(2,3)1 $\otimes$
(5,1)1/2], [(5,0)1 $\otimes$ (2,4)1/2], [(4,2)0 $\otimes$ (3,2)3/2]
$\}$.}
\label{fig9}\end{figure}

The low energy spectra of $^{25}${Mg} is depicted in Fig. \ref{fig7}.
The first column shows the predicted levels, in the second one the
experimental values and in third one those found in full shell model
calculations.
Its band structure is shown in Fig. \ref{fig8} and the percentage each irrep
contributes to the wave functions of the ground state, K = 1/2, $1/2_2$,
9/2, 13/2 and 11/2 bands is presented in Fig. \ref{fig9}. The
Hamiltonian parameters used in this calculation are shown in the last row
of Table 2.

\begin{table}
\begin{tabular}{lc|lc}
$(\lambda_\pi, \mu_\pi )S_\pi$ & $C_2$ & $(\lambda_\nu, \mu_\nu )S_\nu$ &
$C_2$ \\ \hline
(4,2)0  & 46 & (5,1)1/2     & 49 \\
(5,0)1  & 40 & (2,4)1/2     & 46 \\
(2,3)1  & 34 & (3,2)1/2,3/2 & 34 \\
(0,4)0  & 28 & (4,0)1/2     & 28 \\
(3,1)0,1& 25 & (1,3)1/2,3/2 & 25 \\
(1,2)1,2& 16 & (2,1)1/2,3/2 & 16 \\
(2,0)0  & 10 & (0,2)1/2,5/2 & 10 \\
(0,1)1  &  4 & (1,0)3/2     &  4
\label{t8}\end{tabular}
\caption{ Irreps, spins and  $C_2$ values for protons and neutrons in
$^{25}${Mg}.}
\end{table}

\begin{table}
\begin{tabular}{cl|lll}
$(\lambda_\pi, \mu_\pi )$ & $(\lambda_\nu, \mu_\nu )$ &
\multicolumn{3}{c}{$(\lambda, \mu )$ total} \\ \hline
(4,2)0    & (5,1)1/2     & (9,3)1/2 & (10,1)1/2 & (7,4)1/2  \\
(4,2)0    & (2,4)1/2     & (6,6)1/2 & (7,4)1/2 & (4,7)1/2 \\
(5,0)1    & (5,1)1/2     & (10,1)1/2,3/2 \\
(4,2)0    & (3,2)1/2,3/2 & (7,4)1/2,3/2 \\
(5,0)1    & (2,4)1/2     & (7,4)1/2,3/2 \\
(2,3)1    & (5,1)1/2     & (7,4)1/2,3/2 \\
(2,3)1    & (2,4)1/2     & (4,7)1/2,3/2
\label{t9}\end{tabular}
\caption{ The 16 irreps used in description of $^{25}${Mg}.}
\end{table}

The energy levels shown in Fig. \ref{fig7} exhibit a complicated
pattern. To identify the band structure depicted in Fig. \ref{fig8}, both
the B(E2) values and the wave function decomposition were employed. This
band structure is fully consistent with the one reported in
\cite{Col1975}.
While good agreement has been found for a number of states, others like
$9/2_1$, $1/2_2$, $11/2_1$, and $13/2_1$ show a deviation from the
experimental and shell model values, reflecting a limitation of the
present model, probably due both to the Hamiltonian used and the truncation of
the Hilbert space.

The description of the $^{25}${Mg} energy levels and its wave 
functions in terms of
SU(3) irreps reported here is very similar to the one described in 
the original work of
Draayer \cite{Dra73}, despite the schematic interaction and the 
smaller basis used in
the present work. It strongly supports the reliability of the quasi 
SU(3) truncation
scheme.

The general structure of the energy spectra in  Fig. \ref{fig7}
reproduces
some of
the observed levels. The limitations of the schematic
Hamiltonian used can be gauged by the``band shifts" \cite{Col1975} of
three excited bands: those with K = $1/2_2$, 9/2 and 13/2. The effect of
other terms in the Hamiltonian, like proton - neutron pairing, in shifting
these bands in the right direction will be the subject of future research.

In view of the limitations of Hamiltonian (\ref{eq:ham}), a very
specific selection of rotor terms was needed to get a 5/2 ground
state. The parameter set shown in the last line of Table 2 reflects a
subtle balance between the $J^2$, $K^2$ and symmetry terms. The negative
$a$ value pushes bands with larger K down in energy, while the positive
$b$ reduces the moment of inertia, increasing the level separation inside
each band, and moving the band heads with larger $J$ to higher
energies. Setting  $A_{sym} =$ 0.024 helps to make the $J=5/2_1$ state the
ground state.

\begin{table}
  \begin{tabular}{c|cccc}
$ B(E2) \uparrow $ & Exp.  & SU(3) & \cite{Dra73}\\ \hline
$1/2_1^+ \rightarrow 5/2_1^+$ & 0.024 $\pm$ 0.001 & 0.398 &0.530\\
$3/2_1^+ \rightarrow 5/2_1^+$ & 0.033 & 0.276 &0.180\\
$1/2_1^+ \rightarrow 3/2_1^+$ & 0.868 $\pm$ 0.434 & 1.249 &1.440\\
$5/2_1^+ \rightarrow 7/2_1^+$ & 1.621 $\pm$ 0.289 & 1.968 &1.813\\
$3/2_1^+ \rightarrow 5/2_2^+$ & 0.202 $\pm$ 0.098 & 0.835Ý&0.510\\
$1/2_1^+ \rightarrow 5/2_2^+$ & 2.345 $\pm$ 1.042 & 2.323 &2.310\\
$1/2_2^+ \rightarrow 5/2_1^+$ & 0.156 $\pm$ 0.074 & 0.020 &0.044\\
$3/2_1^+ \rightarrow 7/2_2^+$ & 2.084 $\pm$ 0.261 & 1.488 &1.540\\
$5/2_1^+ \rightarrow 7/2_2^+$ & 0.010 & 0.001 &0.001\\
$3/2_2^+ \rightarrow 5/2_1^+$ & 0.074 & 0.002Ý&0.004\\
$7/2_1^+ \rightarrow 9/2_1^+$ & 0.814 $\pm$ 0.271 &0.081*)&1.362\\
$5/2_1^+ \rightarrow 9/2_1^+$ & 0.550 $\pm$ 0.043 & 0.432 &0.400\\
$7/2_1^+ \rightarrow 9/2_2^+$ & 0.054 $\pm$ 0.038 &1.420*)& \\
$5/2_1^+ \rightarrow 9/2_2^+$ & 0.097 $\pm$ 0.006 & 0.545 & \\
$5/2_2^+ \rightarrow 9/2_3^+$ & 2.171 $\pm$ 0.362 & 2.066Ý& \\
$7/2_2^+ \rightarrow 11/2_1^+$ &0.847 $\pm$ 0.261 & 0.001 & \\
$7/2_1^+ \rightarrow 11/2_1^+$ &0.332 $\pm$ 0.078 & 0.372 & \\
$9/2_1^+ \rightarrow 13/2_1^+$ &0.128 $\pm$ 0.061 & 0.279Ý&

\label{t10}
\end{tabular}

\caption{ B(E2) transitions for $^{25}${Mg} in [$e^2b^2 \times 
10^{-2}$]. The second
column shows the experimental values, the third column the present 
theoretical results,
and fourth those obtained previously with the SU(3) model 
\cite{Dra73}. Stars denote
transitions with unclear identification in the theory.}
\end{table}

Table 10 lists the B(E2) transitions for $^{25}${Mg}, calculated with the
same effective charges $e_{\pi ~eff} = 1.56 e, e_{\nu ~eff} = 0.56 e$ 
as used in the
previous cases. The second column shows the experimental values, the 
third column the
present theoretical results, and fourth those obtained previously using
the SU(3) model with a renormalized Kuo-Brown interaction
\cite{Dra73}. The agreement is in general quite good.

Stars were used to mark the calculated transition strengths B(E2; $7/2_1
\rightarrow 9/2_1$) = 0.081 ([$e^2 b^2 \times 10^{-2}$]) and B(E2; $7/2_1
\rightarrow 9/2_2$) = 1.420. Due a shift of the K=9/2 band, the
$9/2_1$ and $9/2_2$ are inverted in our level scheme. A more natural
assignment would be B(E2; $7/2_1 \rightarrow 9/2_1$) = 1.420 and
B(E2; $7/2_1 \rightarrow 9/2_2$) = 0.081, which are close to the
experimental results.


Calculated results for the B(E2) transition strengths are not very 
good. Only a few are
well described:
$1/2_1^+ \rightarrow 3/2_1^+$, $5/2_1^+ \rightarrow 7/2_1^+$, $1/2_1^+
\rightarrow 5/2_2^+$, $5/2_1^+ \rightarrow 9/2_1^+$, $5/2_2^+ \rightarrow
9/2_3^+$, $7/2_1^+ \rightarrow 11/2_1^+$, and $9/2_1^+ \rightarrow
13/2_1^+$. For the other reported transitions the difference is one or more
orders of
  magnitude, most being underestimated. These wrong predictions
show
the limitations of the model. It is interesting to see that the present
values and those reported in Ref. \cite{Dra73} are in general very close.

As mentioned above, by using the B(E2) transition strengths and the 
form of the wave
function it was possible to identify the six rotational bands shown in
Fig.
\ref{fig8}, in agreement with \cite{Col1975}. The SU(3) content of 
these bands is
shown in Fig.
\ref{fig9} as a function of the angular momentum of the states belonging to the
  different
bands. Beyond the regular structure found in most of the bands, the most
  remarkable
feature is that of first two bands which have band heads that are dominated by
the (6,6) irrep. The ``leading" irrep
  (9,3)1/2
is dominant in the second excited K=$1/2_2$ band. This is a 
remarkable result: the
quadrupole - quadrupole interaction builds the ground state band 
mostly from the leading
irrep. As was pointed out in \cite{Dra73}, it
also implies a coexistence of prolate and triaxial shapes, associated 
with the (9,3)
and (6,6) irreps, respectively. It also shows that, while developing 
clear rotational
bands, the $^{25}$Mg ground state is mostly triaxial. The yrast band, however,
includes many different band heads, which are either triaxial or prolate.

At variance from what was found in the lighter nuclei discussed 
above, the spin 3/2
irreps play only a marginal role in $^{25}$Mg, except for the states 
with the largest
angular momentum in the K = 9/2 band, insert d), and the K= 11/2 
band, insert f),
which is dominated by the (7,4) 3/2 irrep.

\section{Discussion and Outlook}

The quasi SU(3) truncation scheme was used in conjunction with Hamiltonian
(\ref{eq:ham}) to describe the energy spectra and B(E2) transition
strengths in $^{21}$Ne, $^{23}$Na and $^{25}$Mg. Comparison were made
with experimental data, with shell model calculations and with previous
SU(3) studies. The agreement was in general good, exhibiting the success
of the model.

The truncation recipe is quite simple. First select three or four 
proton and neutron
SU(3) irreps with largest $C_2$ values, which in general will have spin 0 or 1
for an even number of nucleons and spin 1/2 and 3/2 for odd an number 
of nucleons. Build
from these the coupled proton-neutron SU(3) irreps, and select again 
only a small
number with the largest total $C_2$ values. Usually 20 irreps are enough.
Use this basis
to diagonalize the Hamiltonian. While the effect of the truncation in
$^{21}$Ne was small, for
$^{25}$Mg it implied a two order of magnitude reduction in the basis size. The
composition of the calculated wave functions was found to be very 
similar to the ones
reported in previous SU(3) studies with far larger basis where the 
Kuo interaction was
used \cite{Dra73}.

The Hilbert space built in this way is rich enough to include many 
excited rotational
bands which have a clear counterpart both in the experimental data 
and in full shell
model calculations. The fact that most of these bands have one fairly dominant
SU(3) irrep underscores the strength of the model.  This feature 
leads to a simple
picture in terms of the SU(3) decomposition of the wave functions, 
and exhibits the
crossing and mixings which occur within a band as a function of the 
angular momentum of
the different band members.
B(E2) transition strengths were also found to be in close 
correspondence with the
experimental data, and this allowed for a clear identification of band members.
The SU(3) content of the ground state band also showed that, while 
$^{21}$Ne and
$^{23}$Na are definitively prolate, $^{25}$Mg is mostly triaxial.

There were model limitations uncovered in the present study. The most
stricking are the band shifts, i.e. the fact that some excited bands are
predicted at energies  1 to 2 MeV lower than they appear in the
experiment. Given that the band structure and the B(E2) values were in
general well depicted, the band shifts seem to reflect on a limitation
of the Hamiltonian and not of the Hilbert space,
since the same feature was found in full shell model calculations.

It should be clear that Hamiltonian (\ref{eq:ham}) is not missing important
two-body terms. In addition to realistic single particle energies, it includes
quadrupole-quadrupole and like particle pairing terms with fixed interaction
strengths taken from systematics. It has also three rotor like terms which
allow for a fine tuning of the spectra. In the case of $^{25}$Mg, 
these terms played an
important role in the reproducing of the energy spectra, especially 
in pushing the
$J=5/2$ band down to become the ground state band of the system.

In these nuclei with N very close to Z, proton-neutron pairing, both in
the T=0 and T=1 channels, plays a very important role \cite{Kre74}. Its
inclusion in Hamiltonian (\ref{eq:ham}) would make it isoscalar. It
could be relevant not only for improving the predicted energy spectra, but
also in the description of M1 excitations, which are known to be very
challenging for any theoretical model \cite{Van74}. It would also
allow the model to be tested in the fp-shell where B(M1) transition
strengths have been recently measured \cite{Lis99}. Future research on
these subjects is desirable.

\section{Acknowledgements}

The authors thank C. Johnson and S. Pittel for emphasizing the relevance of
proton-neutron pairing in the $sd$-shell. The authors thank the Institute
of Nuclear Theory of the University of Washington for its hospitality and
the Department of Energy for partial support during the completion of this
work, which was also supported in part by Conacyt (M\'exico) and the
National Science Foundation under Grant PHY-9970769 and Cooperative
Agreement EPS-9720652 that includes matching from the Louisiana Board of
Regents Support Fund.

\end{document}